\documentclass[a4,12pt]{article}
\usepackage[dvips]{graphicx,psfrag}
\usepackage{float}
\usepackage{amsmath,amsthm}
\usepackage{multirow}
\usepackage{arydshln}
\usepackage{cite}

\def\ls{\lower0.5ex\hbox{$\buildrel >\over{\scriptstyle\sim}$}}
\def\rs{\lower0.5ex\hbox{$\buildrel <\over{\scriptstyle\sim}$}} 
\allowdisplaybreaks
\begin{document}
\pagestyle{empty} \setlength{\footskip}{2.0cm}
\setlength{\oddsidemargin}{0.5cm}
\setlength{\evensidemargin}{0.5cm}
\renewcommand{\thepage}{-- \arabic{page} --}
\def\mib#1{\mbox{\boldmath $#1$}}
\def\bra#1{\langle #1 |}  \def\ket#1{|#1\rangle}
\def\vev#1{\langle #1\rangle} \def\dps{\displaystyle}
 \def\thebibliography#1{\centerline{REFERENCES}
 \list{[\arabic{enumi}]}{\settowidth\labelwidth{[#1]}\leftmargin
 \labelwidth\advance\leftmargin\labelsep\usecounter{enumi}}
 \def\newblock{\hskip .11em plus .33em minus -.07em}\sloppy
 \clubpenalty4000\widowpenalty4000\sfcode`\.=1000\relax}\let
 \endthebibliography=\endlist
 \def\sec#1{\addtocounter{section}{1}\section*{\hspace*{-0.72cm}
 \normalsize\bf\arabic{section}.$\;$#1}\vspace*{-0.3cm}}
\def\secnon#1{\section*{\hspace*{-0.72cm}
 \normalsize\bf$\;$#1}\vspace*{-0.3cm}}
 \def\subsec#1{\addtocounter{subsection}{1}\subsection*{\hspace*{-0.4cm}
 \normalsize\bf\arabic{section}.\arabic{subsection}.$\;$#1}\vspace*{-0.3cm}}
\vspace*{-1.7cm}
\renewcommand{\labelitemii}{$\circ$}
\begin{flushright}
$\vcenter{
}$
\end{flushright}

\vskip 1.6cm
\begin{center}
  {\Large \bf Studying flavor-changing neutral $\mib{tqZ}$ couplings$\,$:}
\vskip 0.18cm
  {\Large\bf Current constraints and future prospects}
\end{center}

\vspace{0.5cm}
\begin{center}
\renewcommand{\thefootnote}{\alph{footnote})}
Zenr\=o HIOKI$^{\:1),\:}$\footnote{E-mail address:
\tt hioki@tokushima-u.ac.jp}\ 
Kazumasa OHKUMA$^{\:2),\:}$\footnote{E-mail address:
\tt ohkuma@ice.ous.ac.jp}\ and\
Akira UEJIMA$^{\:2),\:}$\footnote{E-mail address:
\tt uejima@ice.ous.ac.jp}\

\end{center}

\vspace*{0.4cm}
\centerline{\sl $1)$ Institute of Theoretical Physics,\ University of Tokushima}

\centerline{\sl Tokushima 770-8502, Japan}

\vskip 0.2cm
\centerline{\sl $2)$ Department of Information and Computer Engineering,}

\centerline{\sl Okayama University of Science}

\centerline{\sl Okayama 700-0005, Japan}

\vspace*{1.8cm}

\centerline{ABSTRACT}

\vspace*{0.2cm}
\baselineskip=21pt plus 0.1pt minus 0.1pt
Possible non-standard $tuZ$ and $tcZ$ interactions, which induce flavor-changing neutral-current decays
of the top quark, are studied in the effective-Lagrangian framework.
The corresponding Lagrangian consists of four kinds of non-standard couplings
coming from $SU(3)\times SU(2) \times U(1)$ invariant dimension-6 effective operators.
The four coupling constants in each interaction are treated as complex numbers independent of each other,
and constraints on them are derived by using the present 
experimental limits of the branching fractions for $t\to u Z$ and $t\to c Z$ processes.
Future improvements of those constraints are also discussed as well 
as possibilities of measurements of these couplings at 
the High-Luminosity Large Hadron Collider.
\vskip 1.5cm

\vfill

\setcounter{page}{0}
\newpage
\renewcommand{\thefootnote}{$\sharp$\arabic{footnote}}
\pagestyle{plain} 
\setcounter{footnote}{0}

\sec{Introduction}
The Large Hadron Collider (LHC) established the standard model of particle physics 
by discovering the Higgs boson~\cite{Aad:2012tfa, Chatrchyan:2012xdj}, which was the last piece
in the model.
Meanwhile, the LHC has also been searching for new particles
which are not included in the framework of the standard model,
however, any signal indicating their existence 
has not been directly observed yet even at 13 TeV.  
Therefore, precise measurements of various standard-particle processes have received a lot of attention as an indirect search
for  new physics.
In particular, processes caused by a Flavor-Changing Neutral Current (FCNC) are quite attractive as a target for
current and future measurements.
That is, the FCNC is strongly suppressed by the Glashow-Iliopoulos-Maiani mechanism in the standard model~\cite{Glashow:1970gm},
and hence observation of such processes will be good indication of new-physics effects.

On the other hand, the top quark, which is the heaviest in the observed el\-e\-men\-ta\-ry-particle spectrum,
is also expected to play a crucial role in new-physics search 
(see, e.g., \cite{Russell:2017cut} and its references) 
because of the following reasons:\\
(1) The top quark decays without receiving non-perturbative hadronization effects
        due to a very short lifetime originated from its huge mass close to the electroweak scale~\cite{Bigi:1980az, Bigi:1986jk}.
        This unique quality is suitable for future clean analyses of its interactions, which have not been
        studied enough accurately yet in comparison with the other lighter quarks and still might have
        room to hide some non-standard terms.
\\
(2) The $C\!P$-violation in the top-quark sector is known to be very small within the standard model.
        Therefore, if any sizable $C\!P$-violation effects are measured in top-quark productions and/or decays,
        they can be interpreted to come from a possible new physics beyond the standard model.
\\
(3) The top quark strongly interacts with the yet-mysterious Higgs boson.
 	Thus, detailed studies of top-quark interactions would be useful to 
        explore the mechanism of electroweak symmetry breaking as well as properties of the Higgs boson.

Considering them altogether and also taking into account the fact that top-quark precise measurement is one of
the most important missions at future facilities (e.g., High-Luminosity Large Hadron Collider,
International Linear Collider, and Future Circular Collider), we focus in this letter on
flavor-changing neutral-current decays of the top quark $t\to uZ$ and $t\to cZ$, and study to what extent
we can draw valuable information on non-standard interactions that could induce these processes from the present
experimental data. Their branching fractions are estimated to be of ${\cal O}(10^{-17})$ and ${\cal O}(10^{-14})$
respectively in the standard model\cite{AguilarSaavedra:2004wm}, that is why
observation of those processes will be a clear evidence of possible new physics beyond the standard model.

It is not that easy, however, 
to identify what kinds of new interactions had contributed to the processes, even if we observed
some meaningful signals.
In such a situation, the effective-Lagrangian approach which examines model-independently
all the possible extensions of the standard-model interactions is one of the most promising
methods for evaluating possible new-physics contributions.

Indeed, a lot of efforts have already been made to analyze the $tqZ$ interactions, where $q=u/c$, 
based on the effective Lagrangian in order to probe top-quark FCNC
\cite{AguilarSaavedra:2004wm,Han:1995pk,Han:1998yr,delAguila:1999ac,delAguila:1999kfp,Larios:2004mx,Larios:2006pb,
Coimbra:2008qp,Drobnak:2008br,Ferreira:2009bf,Datta:2009zb,Khanpour:2014xla,Durieux:2014xla,
Goldouzian:2014nha,Aguilar-Saavedra:2017vka,Shen:2018mlj}.
In most of the studies, however, only some couplings included in the effective Lagrangian have been
treated as free parameters at once fixing the others.
Furthermore, these non-standard couplings tend to be restricted to real or partially imaginary numbers there,
though they could be complex in general.

Although such limited analyses could be reasonable if the authors are implicitly considering some specific models,
we here deal with all the non-standard couplings as complex numbers and regard 
those couplings as free independent parameters in order to perform a fully model-independent analysis. 
Thus, our results given below do not depend on any specific models.
It never means, of course, that our results offer no
helpful information for model-dependent new-physics studies. That is, they do not have to take into account
the parameter space which has been excluded in our analyses.

The outline of this letter is as follows:
After describing our calculational framework in section 2, 
we perform numerical analyses and show main results in section 3. 
There, we also present some related discussions. 
Finally, a summary is given in section 4.
\sec{Framework}\label{sec:frame}
The effective Lagrangian used here is constructed assuming that there exists some new physics characterized
by an energy scale ${\mit\Lambda}$, and all the non-standard particles are much heavier than the
colliding energies of the LHC.
In this framework, new-physics phenomena are described by $SU(3)\times SU(2) \times U(1)$ invariant
operators whose mass-dimension is six, and there are five kinds of operators that could contribute
to the $tqZ$ interactions~\cite{Buchmuller:1985jz,Arzt:1994gp,AguilarSaavedra:2008zc,Grzadkowski:2010es}.
The appropriate operators are extracted from Ref.~\cite{AguilarSaavedra:2008zc} as follows:
\begin{alignat*}{2}
  {\cal O}^{3,i3}_{\phi q} &=
i \left(\phi^\dagger \tau^I D_{\mu}\phi)
(\bar{q}_{Li}\gamma^{\mu}\tau^I q_{L3}\right),
& \quad  
{\cal O}^{1,i3}_{\phi q} &=
i \left(\phi^\dagger D_{\mu}\phi)
(\bar{q}_{Li}\gamma^{\mu} q_{L3}\right), \nonumber\\
 {\cal O}^{i3}_{\phi u} &=
i \left(\phi^\dagger D_{\mu}\phi)
(\bar{u}_{Ri}\gamma^{\mu} u_{R3}\right),
  & \quad  &   \\
{\cal O}^{i3}_{u W} &=
 \left(
\bar{q}_{Li}\sigma^{\mu\nu}\tau^I u_{R3}
\right)
\tilde{\phi}W^I_{\mu\nu},
& \quad 
{\cal O}^{i3}_{uB\phi} &=
 \left(
\bar{q}_{Li}\sigma^{\mu\nu}u_{R3}
\right)
\tilde{\phi}B_{\mu\nu},
 \nonumber 
\end{alignat*}
where subscripts $i$ and 3 stand for the quark generations,
i.e., $i = 1$ and $2$ correspond to the up and charm quarks, 3 corresponds to the top quark, respectively.

Using the above operators, we can write down the effective Lagrangian of $tqZ$ interactions
describing phenomena around the electroweak scale as
\begin{alignat}{1}\label{eq:efflag_decay}
  &{\cal L}_{tqZ}  = -\frac{g}{2 \cos \theta_W} 
  \Bigl[\,\bar{\psi}_q(x)\gamma^\mu(f_1^L P_L + f_1^R P_R)\psi_t(x)Z_\mu(x) \Bigr.
  \nonumber\\
 &\phantom{========}
  +\bar{\psi}_q(x)\frac{\sigma^{\mu\nu}}{M_Z}(f_2^L P_L + f_2^R P_R)
   \psi_t(x)\partial_\mu Z_\nu(x) \,\Bigr],
\end{alignat}
where  $g$ and $\theta_W$ are the $SU(2)$ coupling constant and the weak mixing angle,
$P_{L/R}\equiv(1\mp\gamma_5)/2$, and $f_{1/2}^{L/R}$ stand for the non-standard couplings defined as
\begin{alignat*}{2}\label{eq:eff_coupling}
f_1^L &=\frac{v^2}{2{\mit\Lambda}^2}\left[
C_{\phi q}^{(3,i3)}+C_{\phi q}^{(3,3i)*}
-C_{\phi q}^{(1,i3)}-C_{\phi q}^{(1,3i)*}
\right],
& \quad 
f_1^R&= -\frac{v^2}{2{\mit\Lambda}^2}\left[
C_{\phi u}^{i3}+C_{\phi u}^{3i*}
\right],\\
f_2^L&=-\frac{\sqrt{2}v^2}{{\mit\Lambda}^2}\left[
C_{uW}^{3i*}\cos\theta_W
-C_{uB\phi}^{3i*}\sin\theta_W
\right],
& \\
f_2^R&=-\frac{\sqrt{2}v^2}{{\mit\Lambda}^2}\left[
C_{uW}^{i3}\cos\theta_W
-C_{uB\phi}^{i3}\sin\theta_W
\right]. & \quad
\end{alignat*}
As mentioned above, $f_{1/2}^{L/R}$ are treated as complex numbers independent
of each other from the viewpoint of model-independent analysis.

\sec{Numerical analyses}\label{sec:r_and_d}
We are now ready to derive allowed regions for the non-standard couplings.
In the analyses, we use the following recent experimental data at 95 \% confidence level:
\begin{quote}
\begin{itemize}
 \item The total-decay width of the top quark, ${\mit\Gamma}^t$ [GeV]~\cite{Aaboud:2017uqq}
 \footnote{
 Two comments on this experimental ${\mit\Gamma}^t$: 
  \begin{quote}
  \begin{description}
    \item[(1)] The actual experimental value presented in \cite{Aaboud:2017uqq} is
    ${\mit\Gamma}^t = 1.76\pm 0.33({\rm stat.})^{+0.79}_{-0.68}({\rm syst.})~{\rm GeV}$.
    However, it is not easy to handle an asymmetric error like this in the error propagation.
    We therefore use ${\mit\Gamma}^t = 1.76\pm 0.33({\rm stat.})\pm 0.79({\rm syst.})~{\rm GeV}$,
    the one symmetrized by adopting the larger (i.e., $+0.79$) in this systematic error.
    
    \item[(2)] 
    There are also measurements of ${\mit\Gamma}^t$ with much smaller errors, 
    but those results were obtained assuming $\sum_q {\rm Br}(t\to Wq)=1$.
    Therefore, it will not be suitable to carry out model-independent analyses based them (see later discussion on this point).
 \end{description}
 \end{quote}
};
\begin{equation}\label{eq:total_w}
4.8 \times 10^{-2} \leq {\mit\Gamma}^t \leq 3.5
\end{equation}
 \item The upper limits of the branching fractions for $t\to qZ$ decays~\cite{ATLAS:2017beb};
\begin{equation}\label{eq:branc}
\begin{split}
 &{\rm Br}(t \to u Z ) < 1.7 \times 10^{-4} \\
 &{\rm Br}(t \to c Z ) < 2.3 \times 10^{-4}
\end{split}
\end{equation}

\end{itemize}
\end{quote}
Then, multiplying the minimum (maximum) value of ${\mit\Gamma}^t$ by ${\rm Br}(t \to u Z/c Z)$,
the corresponding partial decay widths, ${\mit\Gamma_{tqZ}}$ [GeV], are obtained as
\begin{equation}\label{eq:gamma_eff}
\begin{split}
 &0\leq{\mit\Gamma}_{tuZ} < 8.1 \times 10^{-6} ~~(5.9 \times 10^{-4}),\\
 &0\leq{\mit\Gamma}_{tcZ} < 1.1 \times 10^{-5} ~~(8.0 \times 10^{-4}).
\end{split}
\end{equation}


%
In order to find constraints on $f_{1/2}^{L/R}$, 
we compare the experimental limits of ${\mit\Gamma_{tqZ}}$ in Eq.(\ref{eq:gamma_eff})
with the 
one which is calculated from Eq.(\ref{eq:efflag_decay}) as a function of $f_{1/2}^{L/R}$. 
More specifically, by varying the real and imaginary parts of each $f_{1/2}^{L/R}$
independently at the same time,
we look for allowed
regions of them numerically.
In this analysis, 
we take as $m_t=172.5~{\rm GeV}$ and $M_Z=91.2~{\rm  GeV}$
\footnote{Since the masses of the up and charm quarks are much smaller than those of the top quark and the $Z$ boson
and their effects are negligible, we treat these two light quarks to be massless.},
and other physical constants are referred from the Review of Particle Physics ~\cite{Tanabashi:2018oca}.
In addition, we do not neglect any order products of the non-standard couplings, though higher
order products of those couplings tend to be out of consideration in many analyses.
These computations are straightforward, but the parameter space to be explored is quite large, 
so the calculator with Graphics Processing Unit is used to perform them.
%

Table~\ref{tab:tuz_current} shows the obtained constraints on the $tuZ$-coupling parameters,
where those over (under) the dashed lines in the rows marked by Min. and Max. are the minimum and maximum values
of the allowed ranges derived from ${\mit\Gamma}_{tuZ} = 8.1 \times 10^{-6} ~(5.9 \times 10^{-4})$.
Table~\ref{tab:tcz_current} is the same as Table~\ref{tab:tuz_current}
but for the $tcZ$ couplings, where those over (under) the dashed lines are from 
${\mit\Gamma}_{tcZ} = 1.1 \times 10^{-5} $ ($8.0 \times 10^{-4}$).
\begin{table}[tb]
\centering
\caption{Current constraints on the $tuZ$-coupling parameters: Those over (under) the dashed lines
in the rows denoted as Min. and Max. are the minimum and maximum of the allowed ranges
coming from ${\mit\Gamma}_{tuZ} = 8.1 \times 10^{-6} ~(5.9 \times 10^{-4})$.}
\label{tab:tuz_current}
\vspace*{0.3cm}
\begin{tabular}{ccc|cc}
\multicolumn{1}{l}{}                     & \multicolumn{2}{c|}{$f_1^L$}                                                                           & \multicolumn{2}{c}{$f_1^R$}                                                                           \\ \cline{2-5} 
\multicolumn{1}{l}{}                     & Re($f_1^L$)                                       & Im($f_1^L)$                                        & Re($f_1^R$)                                       & Im($f_1^R$)                                       \\ \hline
\multicolumn{1}{c}{\multirow{2}{*}{Min.}}& $-5.5\times 10^{-3}$                              & $-5.5\times 10^{-3}$                               & $-5.5\times 10^{-3}$                              & $-5.5\times 10^{-3}$                              \\ \cdashline{2-5} 
\multicolumn{1}{c}{}                     & \multicolumn{1}{l}{$-4.7\times 10^{-2}$}          & \multicolumn{1}{l|}{$-4.7\times 10^{-2}$}          & \multicolumn{1}{l}{$-4.7\times 10^{-2}$}          & \multicolumn{1}{l}{$-4.7\times 10^{-2}$}          \\ \hline
\multicolumn{1}{c}{\multirow{2}{*}{Max.}}& $\phantom{-}5.5\times 10^{-3}$                    & $\phantom{-}5.5\times 10^{-3}$                     & $\phantom{-}5.5\times 10^{-3}$                    & $\phantom{-}5.5\times 10^{-3}$                    \\ \cdashline{2-5} 
\multicolumn{1}{c}{}                     & \multicolumn{1}{l}{$\phantom{-}4.7\times 10^{-2}$}& \multicolumn{1}{l|}{$\phantom{-}4.7\times 10^{-2}$}& \multicolumn{1}{l}{$\phantom{-}4.7\times 10^{-2}$}& \multicolumn{1}{l}{$\phantom{-}4.7\times 10^{-2}$}\\ \hline
\end{tabular}
\vspace*{0.4cm}
\\
\begin{tabular}{ccc|cc}
\multicolumn{1}{l}{}                     & \multicolumn{2}{c|}{$f_2^L$}                                                                           & \multicolumn{2}{c}{$f_2^R$}                                                                           \\ \cline{2-5} 
\multicolumn{1}{l}{}                     & Re($f_2^L$)                                       & Im($f_2^L)$                                        & Re($f_2^R$)                                       & Im($f_2^R$)                                       \\ \hline
\multicolumn{1}{c}{\multirow{2}{*}{Min.}}& $-4.6\times 10^{-3}$                              & $-4.6\times 10^{-3}$                               & $-4.6\times 10^{-3}$                              & $-4.6\times 10^{-3}$                              \\ \cdashline{2-5} 
\multicolumn{1}{c}{}                     & \multicolumn{1}{l}{$-3.9\times 10^{-2}$}          & \multicolumn{1}{l|}{$-3.9\times 10^{-2}$}          & \multicolumn{1}{l}{$-3.9\times 10^{-2}$}          & \multicolumn{1}{l}{$-3.9\times 10^{-2}$}          \\ \hline
\multicolumn{1}{c}{\multirow{2}{*}{Max.}}& $\phantom{-}4.6\times 10^{-3}$                    & $\phantom{-}4.6\times 10^{-3}$                     & $\phantom{-}4.6\times 10^{-3}$                    & $\phantom{-}4.6\times 10^{-3}$                    \\ \cdashline{2-5} 
\multicolumn{1}{c}{}                     & \multicolumn{1}{l}{$\phantom{-}3.9\times 10^{-2}$}& \multicolumn{1}{l|}{$\phantom{-}3.9\times 10^{-2}$}& \multicolumn{1}{l}{$\phantom{-}3.9\times 10^{-2}$}& \multicolumn{1}{l}{$\phantom{-}3.9\times 10^{-2}$}\\ \hline
\end{tabular}
\vspace*{0.7cm}
\end{table}
\begin{table}[tb]
\centering
\caption{Current constraints on the $tcZ$-coupling parameters: Those over (under) the dashed lines
in the rows denoted as Min. and Max. are the minimum and maximum of the allowed ranges
coming from ${\mit\Gamma}_{tcZ} = 1.1 \times 10^{-5} ~(8.0 \times 10^{-4})$.}
\label{tab:tcz_current}
\vspace*{0.3cm}
\begin{tabular}{ccc|cc}
\multicolumn{1}{l}{}                     & \multicolumn{2}{c|}{$f_1^L$}                                                                           & \multicolumn{2}{c}{$f_1^R$}                                                                           \\ \cline{2-5} 
\multicolumn{1}{l}{}                     & Re($f_1^L$)                                       & Im($f_1^L)$                                        & Re($f_1^R$)                                       & Im($f_1^R$)                                       \\ \hline
\multicolumn{1}{c}{\multirow{2}{*}{Min.}}& $-6.5\times 10^{-3}$                              & $-6.5\times 10^{-3}$                               & $-6.5\times 10^{-3}$                              & $-6.5\times 10^{-3}$                              \\ \cdashline{2-5}
\multicolumn{1}{c}{}                     & \multicolumn{1}{l}{$-5.5\times 10^{-2}$}          & \multicolumn{1}{l|}{$-5.5\times 10^{-2}$}          & \multicolumn{1}{l}{$-5.5\times 10^{-2}$}          & \multicolumn{1}{l}{$-5.5\times 10^{-2}$}          \\ \hline
\multicolumn{1}{c}{\multirow{2}{*}{Max.}}& $\phantom{-}6.5\times 10^{-3}$                    & $\phantom{-}6.5\times 10^{-3}$                     & $\phantom{-}6.5\times 10^{-3}$                    & $\phantom{-}6.5\times 10^{-3}$                    \\ \cdashline{2-5} 
\multicolumn{1}{c}{}                     & \multicolumn{1}{l}{$\phantom{-}5.5\times 10^{-2}$}& \multicolumn{1}{l|}{$\phantom{-}5.5\times 10^{-2}$}& \multicolumn{1}{l}{$\phantom{-}5.5\times 10^{-2}$}& \multicolumn{1}{l}{$\phantom{-}5.5\times 10^{-2}$}\\ \hline
\end{tabular}
\vspace*{0.2cm}
\\
\begin{tabular}{ccc|cc}
\multicolumn{1}{l}{}                     & \multicolumn{2}{c|}{$f_2^L$}                                                                           & \multicolumn{2}{c}{$f_2^R$}                                                                           \\ \cline{2-5} 
\multicolumn{1}{l}{}                     & Re($f_2^L$)                                       & Im($f_2^L)$                                        & Re($f_2^R$)                                       & Im($f_2^R$)                                       \\ \hline
\multicolumn{1}{c}{\multirow{2}{*}{Min.}}& $-5.3\times 10^{-3}$                              & $-5.3\times 10^{-3}$                               & $-5.3\times 10^{-3}$                              & $-5.3\times 10^{-3}$                              \\ \cdashline{2-5} 
\multicolumn{1}{c}{}                     & \multicolumn{1}{l}{$-4.5\times 10^{-2}$}          & \multicolumn{1}{l|}{$-4.5\times 10^{-2}$}          & \multicolumn{1}{l}{$-4.5\times 10^{-2}$}          & \multicolumn{1}{l}{$-4.5\times 10^{-2}$}          \\ \hline
\multicolumn{1}{c}{\multirow{2}{*}{Max.}}& $\phantom{-}5.3\times 10^{-3}$                    & $\phantom{-}5.3\times 10^{-3}$                     & $\phantom{-}5.3\times 10^{-3}$                    & $\phantom{-}5.3\times 10^{-3}$                    \\ \cdashline{2-5} 
\multicolumn{1}{c}{}                     & \multicolumn{1}{l}{$\phantom{-}4.5\times 10^{-2}$}& \multicolumn{1}{l|}{$\phantom{-}4.5\times 10^{-2}$}& \multicolumn{1}{l}{$\phantom{-}4.5\times 10^{-2}$}& \multicolumn{1}{l}{$\phantom{-}4.5\times 10^{-2}$}\\ \hline
\end{tabular}
\vspace*{0.7cm}
\end{table}
%

Comparing these two Tables, 
we find that
the $tuZ$ couplings are more strongly restricted than the $tcZ$ couplings. Apparently it comes
from the difference between the current experimental limits of each branching fraction.
On the other hand, both the real and imaginary parts of $f_{1}^{L/R}$ and $f_2^{L/R}$ 
in each of the $tuZ$ and $tcZ$ couplings have the same minimum and maximum limits, respectively.

\begin{table}[tb]
\centering
\vspace*{0.4cm}
\caption{Allowed minimum and maximum values of the $tuZ$ couplings for ${\mit\Gamma}_{tuZ} = 8.1 \times 10^{-6}$
in the case that Re($f_1^L$) is fixed to $5.5\times 10^{-3}$.}
\label{tab:one_fix_sample}
\vspace*{0.3cm}
\begin{tabular}{ccc|cc}
\multicolumn{1}{l}{}     & \multicolumn{2}{c|}{$f_1^L$}                                          & \multicolumn{2}{c}{$f_1^R$}                                      \\ \cline{2-5} 
\multicolumn{1}{l}{}     & Re($f_1^L$)                          & Im($f_1^L)$                    & Re($f_1^R$)                    & Im($f_1^R$)                     \\ \hline
\multicolumn{1}{c}{Min.} & \multirow{2}{*}{$\,\;5.5\times 10^{-3}\;$} & $-1.0\times 10^{-3}$           & $-1.0\times 10^{-3}$           & $-1.0\times 10^{-3}$            \\ \cline{1-1}\cline{3-5} 
\multicolumn{1}{c}{Max.} &  {\scriptsize (Fixed)}               & $\phantom{-}1.0\times 10^{-3}$ & $\phantom{-}1.0\times 10^{-3}$ & $\phantom{-}1.0\times 10^{-3}$  \\ \hline
\end{tabular}
\vspace*{0.2cm}
\\
\begin{tabular}{ccc|cc}
\multicolumn{1}{l}{}     & \multicolumn{2}{c|}{$f_2^L$}                                          & \multicolumn{2}{c}{$f_2^R$}                                      \\ \cline{2-5} 
\multicolumn{1}{l}{}     & Re($f_2^L$)                          & Im($f_2^L)$                    & Re($f_2^R$)                    & Im($f_2^R$)                     \\ \hline
\multicolumn{1}{c}{Min.} & $-8.0\times 10^{-4}$                 & $-8.0\times 10^{-4}$           & $-4.2\times 10^{-3}$           & $-8.0\times 10^{-3}$            \\ \hline
\multicolumn{1}{c}{Max.} & $\phantom{-}8.0\times 10^{-4}$       & $\phantom{-}8.0\times 10^{-4}$ & $-3.4\times 10^{-3}$           & $\phantom{-}8.0\times 10^{-3}$  \\ \hline
\end{tabular}
\vspace*{0.1cm}
\end{table}

We here should comment about the results in the tables. What the {\it allowed} ranges express is:
if we give one parameter a value outside its allowed range, we can no longer reproduce the width
${\mit\Gamma}_{tuZ}$ or ${\mit\Gamma}_{tcZ}$ which satisfies the inequalities in Eq.(\ref{eq:gamma_eff})
however we vary the other parameters.
Therefore they do not ensure that any combination of parameter values within the allowed areas
{\it is allowed by} those inequalities.
For example, when one of the parameters is assumed to take its maximum value listed in the tables
(e.g., ${\rm Re}(f_1^L)=5.5\times 10^{-3}$),
we obtain more restrictive results for the remaining ones as shown in Table \ref{tab:one_fix_sample}.
This means that the allowed regions shown in Tables \ref{tab:tuz_current} and \ref{tab:tcz_current} 
are larger than the results of limited analyses in which some of the parameters are
fixed.

As a further study towards 
the High-Luminosity LHC (HL-LHC), 
let us perform a similar analysis
assuming that the branching fractions will be improved to be
${\rm Br}(t \to u Z ) < 8.5 \times 10^{-5}$ 
and 
${\rm Br}(t \to c Z ) < 1.2 \times 10^{-4}$, i.e., 50 \% reduction\footnote{
    Those assumed values correspond to ${\rm Br}(t \to q Z ) < 2.0 \times 10^{-4}(=0.02\%)$, where $q=u+c$.
    They are never unrealistic because CMS collaboration estimates the upper limits of ${\rm Br}(t \to q Z )$
    at 95\% confidence level as 0.01\% with an integrated luminosity of 3000 fb$^{-1}$~\cite{CMS-PAS-FTR-13-016}.}.
Combining them with Eq.(\ref{eq:total_w}), the partial decay widths are
\begin{equation}\label{eq:gamma_eff_expected}
\begin{split}
 &0\leq{\mit\Gamma}_{tuZ} < 4.1 \times 10^{-6} ~~(3.0 \times 10^{-4}),\\
 &0\leq{\mit\Gamma}_{tcZ} < 5.5 \times 10^{-6} ~~(4.0 \times 10^{-4}).
\end{split}
\end{equation}
The results thereby for $t\to uZ$ and $t\to cZ$
are estimated as follows:
 \begin{itemize}
 \item $tuZ$ couplings
	 \begin{equation}
 	\label{eq:tuZ-future}
	\begin{split}
  	\left|{\rm Re/Im}(f_{1}^{L/R})\right|\leq 3.9\times 10^{-3},~~
  	\left|{\rm Re/Im}(f_{2}^{L/R})\right|\leq 3.2\times 10^{-3}\\
 	\left(\left|{\rm Re/Im}(f_{1}^{L/R})\right|\leq 3.3\times 10^{-2},~~
 	\left|{\rm Re/Im}(f_{2}^{L/R})\right|\leq 2.7\times 10^{-2}\right)
 	\end{split}
	\end{equation}
 \item $tcZ$ couplings
 	\begin{equation}
        \label{eq:tcZ-future}
   	\begin{split}
        \left|{\rm Re/Im}(f_{1}^{L/R})\right|\leq 4.6\times 10^{-3},~~
  	\left|{\rm Re/Im}(f_{2}^{L/R})\right|\leq 3.8\times 10^{-3}\\
  	\left(\left|{\rm Re/Im}(f_{1}^{L/R})\right|\leq 3.9\times 10^{-2},~~
  	\left|{\rm Re/Im}(f_{2}^{L/R})\right|\leq 3.2\times 10^{-2}\right) 
	\end{split}
	\end{equation}
\end{itemize}
%
Comparing them with those in Tables~\ref{tab:tuz_current} and \ref{tab:tcz_current}, 
we find that the allowed regions are expected
to be narrowed by about 30 \%  if the assumed branch fractions are realized
at the 
HL-LHC 
(see also Figure \ref{fig:arrowed}).
Therefore, if there existed some new physics which gives the parameters close to the minimum or maximum values in
Tables \ref{tab:tuz_current} and \ref{tab:tcz_current},
FCNC on the top-quark sector could be observed at the  
HL-LHC.

Finally, let us get back to the problem in which one of our input data, i.e. ${\mit\Gamma}^t$, allowed
very large/small values in comparison with the standard-model prediction ${\mit\Gamma}^t=1.322$ GeV~\cite{Gao:2012ja}.
For those who find such ${\mit\Gamma}^t$ unrealistic, we also performed the same analysis but with
this standard-model value instead of the experimental one, $4.8 \times 10^{-2} \leq {\mit\Gamma}^t \leq 3.5$,
which decreases the uncertainty though the results
become less model-independent. The resultant constraints on the non-standard couplings are as follows:

\newpage

\begin{figure}[H]
\begin{center}
\caption{Current and expected constraints on the $tuZ$ and $tcZ$ couplings:
The solid (dashed) lines mean the current (expected) allowed ranges for $tuZ$ and $tcZ$ 
shown in Tables~\ref{tab:tuz_current} and \ref{tab:tcz_current}
(
 Eqs.(\ref{eq:tuZ-future}) and (\ref{eq:tcZ-future})
).
The region between the two inside bars on each line is derived using 
${\mit\Gamma}_{tuZ} = 8.1 \times 10^{-6} (4.1 \times 10^{-6})$,
${\mit\Gamma}_{tcZ} = 1.1 \times 10^{-5} (5.5 \times 10^{-6})$  
and that between the two outside bars is derived using 
${\mit\Gamma}_{tuZ}=  5.9 \times 10^{-4} (3.0 \times 10^{-4})$,
${\mit\Gamma}_{tcZ} = 8.0 \times 10^{-4} (4.0 \times 10^{-4})$.}
\label{fig:arrowed}
\vspace*{0.6cm}
\includegraphics[scale=1.25]{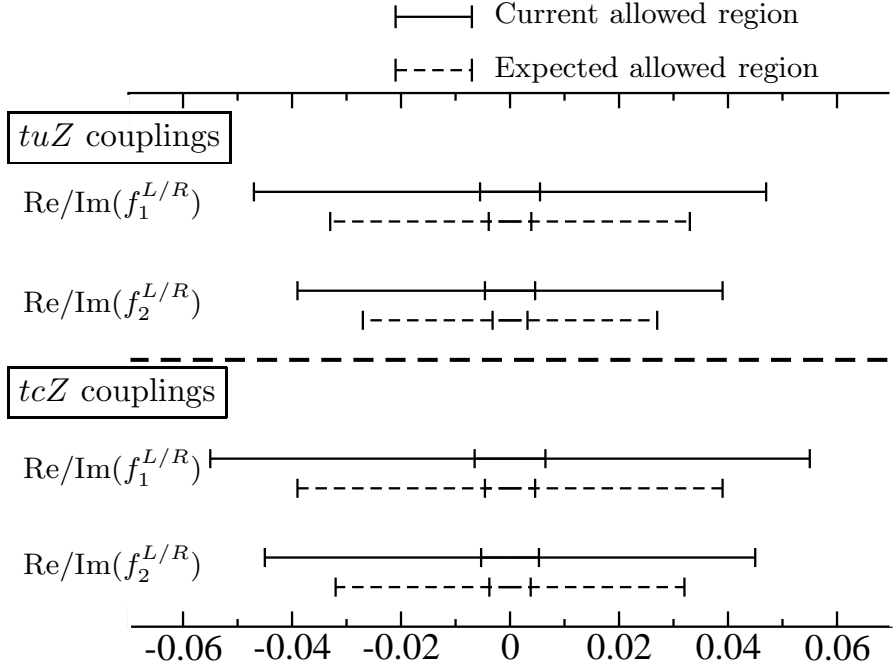}
\end{center}
\vspace*{0.7cm}
\end{figure}

\begin{itemize}
 \item Current constraints (compare with Tables \ref{tab:tuz_current} and \ref{tab:tcz_current})
  \begin{itemize}
 \item $tuZ$ couplings
 $$
  \left|{\rm Re/Im}(f_{1}^{L/R})\right|\leq 2.9\times 10^{-2},~~
  \left|{\rm Re/Im}(f_{2}^{L/R})\right|\leq 2.4\times 10^{-2}
 $$ 
 \item $tcZ$ couplings
 $$
  \left|{\rm Re/Im}(f_{1}^{L/R})\right|\leq 3.4\times 10^{-2},~~
  \left|{\rm Re/Im}(f_{2}^{L/R})\right|\leq 2.8\times 10^{-2}
 $$  
\end{itemize}
 \item Expected constraints (compare with Eqs.(\ref{eq:tuZ-future}) and (\ref{eq:tcZ-future}))
 \begin{itemize}
 \item $tuZ$ couplings
 $$
  \left|{\rm Re/Im}(f_{1}^{L/R})\right|\leq 2.0\times 10^{-2},~~
  \left|{\rm Re/Im}(f_{2}^{L/R})\right|\leq 1.7\times 10^{-2}
 $$ 
 \item $tcZ$ couplings
 $$
  \left|{\rm Re/Im}(f_{1}^{L/R})\right|\leq 2.4\times 10^{-2},~~
  \left|{\rm Re/Im}(f_{2}^{L/R})\right|\leq 2.0\times 10^{-2}
 $$  
\end{itemize}
\end{itemize}
Thus, even if the present experimental ${\mit\Gamma}^t$
is replaced with the standard-model prediction and the uncertainty is neglected,
rather large non-standard coupling constants still could exist.
Note however that two (or more) non-standard couplings among the rest are needed
to take large values at the same time in this case in order to balance out the contribution
from each non-standard coupling.

\sec{Summary}\label{sec:sum}
We have here provided allowed regions of the non-standard $tuZ$ and $tcZ$
couplings via the recent experimental limits of Br($t\to u Z$) and Br($t\to c Z$) 
based on the effective-Lagrangian approach.
In this analysis, we treated all the non-standard couplings as complex numbers
which can vary independently of each other
and gave constraints on them.
It was found that the current allowed regions of these coupling are relatively large
 even if 
we use the standard-model ${\mit\Gamma}^t$ and neglect its uncertainty
because their contributions cancel out each other to a certain extent.
It was also pointed out that the allowed regions derived here
could get narrowed by about 30 \% at the HL-LHC.
This means that there are good chances of discovering
some evidence of the FCNC on the top-quark sector at the HL-LHC,
if some of the current minimum or maximum values of the non-standard couplings are close to 
the actual ones in nature.

\vskip 0.7cm
%
\secnon{Acknowledgments}
%
This work was partly supported by the Grant-in-Aid for Scientific Research (C) 
Grant Number 17K05426 from the Japan Society for the Promotion of Science.
Part of the algebraic and numerical calculations
were carried out on the computer system at Yukawa Institute for
Theoretical Physics (YITP), Kyoto University.

\newpage

\baselineskip=20pt plus 0.1pt minus 0.1pt


\end{document}